\documentclass[12pt]{article}
\usepackage{verbatim,color,amssymb,epsfig}
\usepackage{amsmath}
\begin{document}


\begin{center}
{\LARGE{\bf COVID-19: Should We Test Everyone?} }
\end{center}

\begin{center}

Grace Y. Yi\footnote{\baselineskip=10pt Department of Statistical and Actuarial Sciences,
Department of Computer Science, University of Western Ontario, London, Ontario, Canada, gyi5@uwo.ca}

Wenqing He\footnote{\baselineskip=10pt Department of Statistical and Actuarial Sciences,
University of Western Ontario, London, Ontario, Canada, whe@stats.uwo.ca}

Dennis K. J. Lin\footnote{\baselineskip=10pt Department of Statistics, Pennsylvania State University, University Part, PA, USA,
DKL5@psu.edu}

Chun Ming Yu\footnote{\baselineskip=10pt  Northfield FHO @ Boardwalk Medical Center, Waterloo, Ontario, Canada  N2T 0C1,
cyu245@gmail.com}
\end{center}

\bigskip


\begin{center}
{\Large{\bf Abstract}}
\end{center}

Since the beginning of 2020, the coronavirus disease 2019 (COVID-19) has spread rapidly in the city of Wuhan, P.R. China,
and subsequently, across the world.
The swift spread of the virus is largely attributed to its stealth transmissions in which infected patients may be asymptomatic  or  exhibit only flu-like symptoms in the early stage.
Undetected transmissions present a remarkable challenge for the containment of the virus and
pose an appalling threat to the public. An urgent question that has been asked
by the public is  ``Should I be tested for COVID-19 if I am sick?".
While different regions established their own criteria for screening to identify infected cases, the screening criteria have been  modified
based on
 new evidence and understanding of the virus as well as the  availability of  resources.
The shortage of test kits and medical personnel has considerably limited our ability to
do as many tests as possible.
Public health officials and clinicians are facing a dilemma of balancing the limited resources and
unlimited demands. On one hand, they are striving to achieve the best outcome by
optimizing the usage of the scant resources. On the other hand, they are
challenged by the patients' frustrations and anxieties,
 stemming from the concerns of not being tested for COVID-19 for not meeting the definition of PUI (person under investigation).
In this paper, we evaluate the situation from the statistical viewpoint by
factoring into the considerations of the uncertainty and inaccuracy of the test, an issue that is often overlooked by the general public. We aim to shed light on the tough situation by providing evidence-based reasoning from the statistical angle, and we expect this examination will help the general public understand and assess the situation rationally. Most importantly, the development offers  recommendations for physicians
to make sensible evaluations to optimally use the   limited resources for
 the best medical outcome.

\vspace{8mm}

\par\vfill\noindent

\par\medskip\noindent

\clearpage\pagebreak\newpage

\pagenumbering{arabic}
\section{Introduction} \label{i}



The first case of the coronavirus disease 2019 (COVID-19) was found in December of 2019 in Wuhan city, Hubei providence, P. R. China. On December 31, 2019, China informed the World Health Organization (WHO) of a case of novel viral pneumonia in Wuhan (Wong et al. 2020).
Since the diagnosis of the first case, this virus has spread with astonishing speed and has caused many infections  and a good number of deaths. The origin of the virus, however, remains unclear.
An abrupt announcement was made
on January 23, 2020 that the city of Wuhan was
locked down to control the spread of the virus. Subsequently, almost all  areas in China have begun to take
serious public health measures to
 contain the virus (Xiao and Torok 2020).\\

On January 30, 2020, WHO declared COVID-19 as a public health emergency of international concern.
By March 2, 2020, China had confirmed 80,174 infected cases and 2915 deaths.
From February 27 to March 11, 2020, the number of cases of COVID-19 outside China has increased 10-fold
and the number of affected countries has increased to be 113.
As of March 11, 2020, there have 118,429 confirmed cases and 4,292 deaths (WHO, Situation Reports 2020).
The number of cases, the number of deaths, and the number
 of affected countries are expected to  climb in the coming days and weeks.
 On March 11, 2020, WHO declared  COVID-19 to be a pandemic.\\
%
%

COVID-19 has been found to be caused by the severe acute respiratory syndrome coronavirus 2 (SARS-CoV-2);
 certain epidemiological and
clinical characteristics of patients with COVID-19 have been reported. However, comprehensive knowledge of COVID-19 still remains
incomplete. For instance,
the risk factors for mortality and the
clinical course of the illness
%
%
have not been well understood (Zhou et al. 2020).
The early presentation of COVID-19 infection is typically nonspecific.
Some infected cases
may be asymptomatic, while many infected individuals
 often show flu-like symptoms such as dry
cough, sore throat, low-grade fever, or malaise in the first few days (Wong et al. 2020).
 The seeming-flu symptoms have created difficulties
in  differentiating COVID-19 from the common cold and  seasonal influenza.\\

The mystery of the virus and the lack of  effective treatment for COVID-19  have presented a striking threat to the public.
In contrast to the rapid transmission of
 the COVID-19 pandemic,
 news on COVID-19 has  traveled swiftly and broadly through internet,
 radio, newspaper, television,
 social media, and so on.
 The wave of fear
   has escalated in the public.
Though it has been reported that people with medical complications are at a greater risk of suffering from COVID-19,
 the general public  also has the fear of contracting the virus.\\

Clinicians have been under tremendous  pressure to triage the high volume of
patients for COVID-19 testing.
To
receive  medical care as early as possible, an urgent
question that has  been asked  by
patients with flu-like symptoms is ``Can I be tested for COVID-19?"
More broadly,
the public may be puzzled by the concern,
``To  achieve an effective
containment of the virus,
why does the government not take a proactive action to test
everyone for COVID-19 to identify all infected individuals in a timely manner?"\\

Due to the limited availability of trained personnel, testing kits, and PPE (personal protective
equipment),
 it is impossible to test everyone with flu-like symptoms for COVID-19,
let alone to test every individual. While these reasons can  easily be perceived by the
public, the underlying scientific reasonings do not seem to be  considered. More importantly,
due to the limited availability of resources,
the protocol of screening patients for testing COVID-19 can be stringent to prevent
a collapse in medical facilities.
Based on the evolving global situation, the
 screening process is  constantly being updated and differs  from country to country.
Initially, the U.S. Centers for Disease Control and Prevention (CDC)
 recommended testing only people with respiratory symptoms
 such as fever, dry cough,  shortness of breath,
and those who had potentially been exposed to the virus.
With  the evidence for community transmissions,
  the CDC updated its recommendations on March 4, 2020 to allow
anyone with respiratory symptoms to be
tested as long as the request is approved by  a doctor, though the CDC is encouraging physicians to minimize unnecessary testing by considering patients' exposure risks (Ferran 2020).\\

From the medical perspective, testing for COVID-19 has crucial  implications and importance.
It is impossible to fight the virus blindly without knowing the target population. The early
diagnosis of infected patients is essential to manage the situation.
Infected people must be
 isolated to control the virus spread;  potentially infected individuals should be
quarantined to minimize the possibility of infecting healthy people;
 and vulnerable people such as the elderly
 and patients with chronic health issues need to be
 secluded to prevent infection.
Necessary medical attention must be focused on patients at high risk
who require immediate medical intervention to prevent mortality.
In a broader spectrum of learning and dealing with the virus, acquiring  accurate data of infection and transmission
is critical for
 researchers
to unveil the correct profile of COVID-19 to implement more effective clinical
management.\\

In response to the increasing need for testing for COVID-19, the U.S. Food and Drug Administration (FDA) announced on February 29, 2020, a new policy that made it easier for commercial and academic laboratories to develop their own tests and allowed other certified labs to test patient samples.
Companies, hospitals and  institutions are now racing to develop more tests to diagnose  COVID-19.
On March 10, 2020, Alex Azar, secretary of Health and Human Services, announced that 2.1 million testing kits were available and more than 1 million have shipped to certified labs for testing (Ferran 2020).\\

With the urgency of identifying infected cases and
  increase in available test kits,  screening criteria for  testing COVID-19 have become less restrictive than at the initial stage.
It now seems quite tempting to take aggressive action to adminster  COVID-19 tests to  as many patients as possible.
However, an important yet overlooked issue is on the imperfectness of test procedures.
It is imperative
to  enhance our understanding of testing for COVID-19
by factoring in the assessment of the uncertainty, randomness and
 imperfectness associated with the test procedures; otherwise, misleading and erroneous outcomes can be produced.\\

In this article,
we examine the concerns of testing COVID-19 from the statistical standpoint. Our
explorations  are purely based on accounting for the uncertainty and randomness
associated with medical  test procedures. We will look at
 the uncertainty induced from the test procedures
and  assess the degree of the resulting false results.
Our explorations are intended to shed light on the question ``Should everyone be tested for COVID-19?",
which would assist general people to
 assess situations with rational and evidence-based thinking.
Ultimately, as advocated by Sharkawy (2020), the public
should face the challenge of COVID-19
 with educated reasoning and compassion for others. Everyone should
 seek truth and facts, as opposed to conjecture and speculation; we all must work together to battle COVID-19.\\

This article provides a dynamic framework to present the evolving features of COVID-19.
We  examine
test procedures in terms of their sensitivity and specificity, and  quantify the degrees of false test results
under various scenarios.
Most importantly, we make sensible recommendations for physicians to balance the usage of limited test kits
and the accuracy of the test outcome. We offer
the assessment as to how likely we may miss identifying COVID-19 carriers based on consecutive negative results and how many times we should test a suspected COVID-19 patient to reduce the chance of errors. Our discussion provides the guidelines for discharging patients who are treated as COVID-19 carriers.\\

\section{Notations and Framework}\label{naf}

For generality, we use the term {\em population} to describe the group of subjects of our interest.
In the following discussion, {\em population} may represent the collection of all people in a country, a city, or a region; it may also refer to a cohort of individuals, a ward of patients, or a community of people.
We first introduce abstract symbols to
represent the quantities of our interest.
To facilitate the dynamic feature of
COVID-19, we include the dependence on time in the notations.\\

 On day $t$ with $t= 1,2, \ldots$,
suppose our target population has $N(t)$ people in total in which
  $N_h(t)$ people have no COVID-19  and
$N_s(t)$ people have COVID-19. Let $P(t)=N_s(t)/N(t)$ be the {\em prevalence} on day $t$.
Before  {\em  patient zero} (i.e., the first person who has COVID-19) appears, no one in the population
has the virus. So we assume that
when $t = 0, N_s(t) = 0$,  and when $t=1, 2, \ldots$, $N_s(t) \ge 1$.
That is, $P(t)=0$ for $t=0$ and $0 < P(t) \le 1$ for  $t=1, 2, \ldots$.
Due to the dynamic feature and the spread of the virus,
the relative size of $N_h(t)$ and $N_s(t)$ varies with time $t$.
Initially,   $N_s(t)$ is negligible
and
$N_s(t) \ll N_h(t)$ for $t$ in a certain interval, say $[1, T_1]$, i.e., $P(t)$ is near 0.
As  outbreaks occur and the pandemic is declared, it is possible that
$N_s(t) \approx N_h(t)$ for $t$ in a certain
interval $[T_1, T_2]$, say,
yielding $P(t) \approx 1/2$. In the worst scenario,
 $N_s(t) \gg N_h(t)$ for $t$ in a certain time period $[T_2, T_3]$, say, leading to $P(t) \approx 1$.
Eventually, we hope that
 the state of
coming back to $N_s(t) \approx 0$ or $P(t) \approx 0$ for $t$ in the interval $[T_3, \infty)$ will be reached
with $T_3$ being as small as possible.\\

For {\em any} individual in the population, we are interested in the COVID-19 status for this individual. Let
$Y$ be the binary variable showing  the {\em true} status for an individual to have COVID-19, with
$Y= 1$ if having COVID-19 and 0 otherwise.
In reality, the true value of $Y$ is {\em unknown} for any individual, and we can only
apply proper medical tests to find out an individual's disease status. To feature this,
let  $Y^*$ represent the test result for an individual who is tested;
$Y^* =1$, if the test result is positive; and $Y^*=0$, if the test result is negative.\\

However,  no medical test
is 100\% accurate in practice. There is a chance that a medical test can give us an incorrect result. To describe the accuracy of a test, we use two useful measures, called the {\em sensitivity}
and the {\em specificity}, which are respectively defined as
$$
p_{sen} = P(Y^* = 1 |Y= 1) \ \mbox{and} \ p_{spe} = P(Y^* = 0 | Y= 0).
$$

\vspace{2mm}

Basically, the  sensitivity $p_{sen}$ is a measure to show how sensitive the test is to testing diseased subjects. It reports the probability that
a test successfully confirms the true  status for an individual having the disease. In other words,
the value of $p_{sen}$ indicates the success rate of the test when applied to the {\em subpopulation}
of diseased people, so  the  sensitivity $p_{sen}$ is also called the
{\em true positive rate}. An accurate test is expected to have a value near 1. \\

Paying attention to the  sensitivity $p_{sen}$ only is, however, not enough to characterize the goodness of a test. A good test should also be accurate in terms of correctly showing the result for people who {\em do not} have the disease. To this end, the
specificity $p_{spe}$ comes into the play. It measures  the probability that
the test successfully
 reveals the disease-free status  for any individual who has no disease. The value of $p_{spe}$ indicates the proportion of the time for obtaining the correct result
when
 the test is applied  to the {\em subpopulation}
of healthy people. Consequently,   the  specificity  $p_{spe}$ is also called the
{\em true negative rate}. A good test is also expected to
have  $p_{spe}$ close to 1 to keep the number of misdiagnosed cases small.\\

Corresponding to the true positive rate and the true negative rate, the complement probabilities
$1 - p_{sen} = P(Y^* = 0 |Y=1)$ and $1- p_{spe} = P(Y^* =1 |Y=0)$  are also useful to describe
the test outcomes. These two measures are called the
{\em false negative rate} and the {\em false positive rate}, respectively.\\

Because medical tests are not always accurate in testing diseased or non-diseased individuals, it is important
to take into account the uncertainty and randomness when interpreting a test result. To make a sensible decision, it is necessary to
understand how to evaluate the risk of receiving a false result when applying a test. Ultimately, our goal is to make an educational and evidence-based decision for the health care.
Specifically, we are interested in
evaluating two numbers of
concern,
$$
{\rm \#_{fp}}(t) = \mbox{the number of false positive on day} \ t \ \mbox{if everyone in the population is tested},
$$
and
$$
{\rm \#_{fn}}(t) = \mbox{the number of false negative on day} \ t \ \mbox{if everyone in the population is tested}.
$$

It is easily seen that those numbers are determined by the size of the diseased subpopulation
and the size of the non-diseased subpopulation as well as the sensitivity and the specificity of the test. That is,
\begin{equation}
{\rm \#_{fp}}(t) = N_h(t) \times (1 - p_{spe} ) \ \ \mbox{and} \ \
{\rm \#_{fn}}(t) = N_s(t) \times (1 - p_{sen} ).
\label{false-number}
\end{equation}

\section{Should Everyone  Be Tested?}

To understand how medical tests with different sensitivities and specificities may perform,  we
consider the scenario where the population  has  possibly different prevalence for different days. We make
recommendations by examining how the number of false negative
${\rm \#_{fn}}(t)$ and the number of false positive ${\rm \#_{fp}}(t)$ are determined by the sensitivity and specificity of the test procedure as well as the prevalence of the disease.

\subsection{Tracking Patient Zero}

To visualize the relationship between the false negative number ${\rm \#_{fn}}(t)$ and
 the sensitivity of the tests,
 we start with an example with $N(t)=10,000$ for a given day $t$ and apply a sequence
of tests with different sensitivities to everyone in the population.
We
 show the results
in Figure 1 for  a range of prevalence. As expected,  the false negative number
drops as the sensitivity of the test becomes higher, and the drop rate is higher for the population with
a bigger prevalence than that with a smaller prevalence. When the prevalence is very small, say, $P(t)=10^{-4}$, or equivalently, $P(t)=1/N(t)$ here (i.e., when {\em patient zero} just presents in the population), the false negative number is below 1 no matter what the sensitivity of a test is. In this case, testing everyone for COVID-19 would not virtually yield any false negative result.\\

In general, in the very beginning of the presence of {\em patient zero} (i.e., at time $t=1$), the population has  $N_s(1)$ to be 1 or nearly 1. If everyone is tested for COVID-19, then
$$
{\rm \#_{fn}}(1) = N_s(1) \times (1 - p_{sen} ) < 1
$$
regardless of the accuracy of test procedures.
To track the origin of COVID-19 for a population,
we recommend to test everyone in a
group of presumptive patients who may potentially include {\em patient zero};
this is the only way to identify
{\em patient zero}, yet it is unnecessary to worry about obtaining false negative results,
no matter how inaccurate the test procedure could be.\\

In reality, it is often difficult to immediately identify the presence time of {\em patient zero} in a cohort based on the confirmation of infected cases.
The discussion here offers a possible way to track {\em patient zero} retrospectively by examining the samples of
suspected patients.  Checking the sample of {\em every} suspected patient by the reverse time order is needed to identify {\em patient zero} in the cohort.  \\

\noindent
{\bf Recommendation 1:} {\em To identify patient zero,  it is recommended to test everyone in a presumptive group which may potentially include patient zero.}\\

\begin{center}
\includegraphics[width=3in]{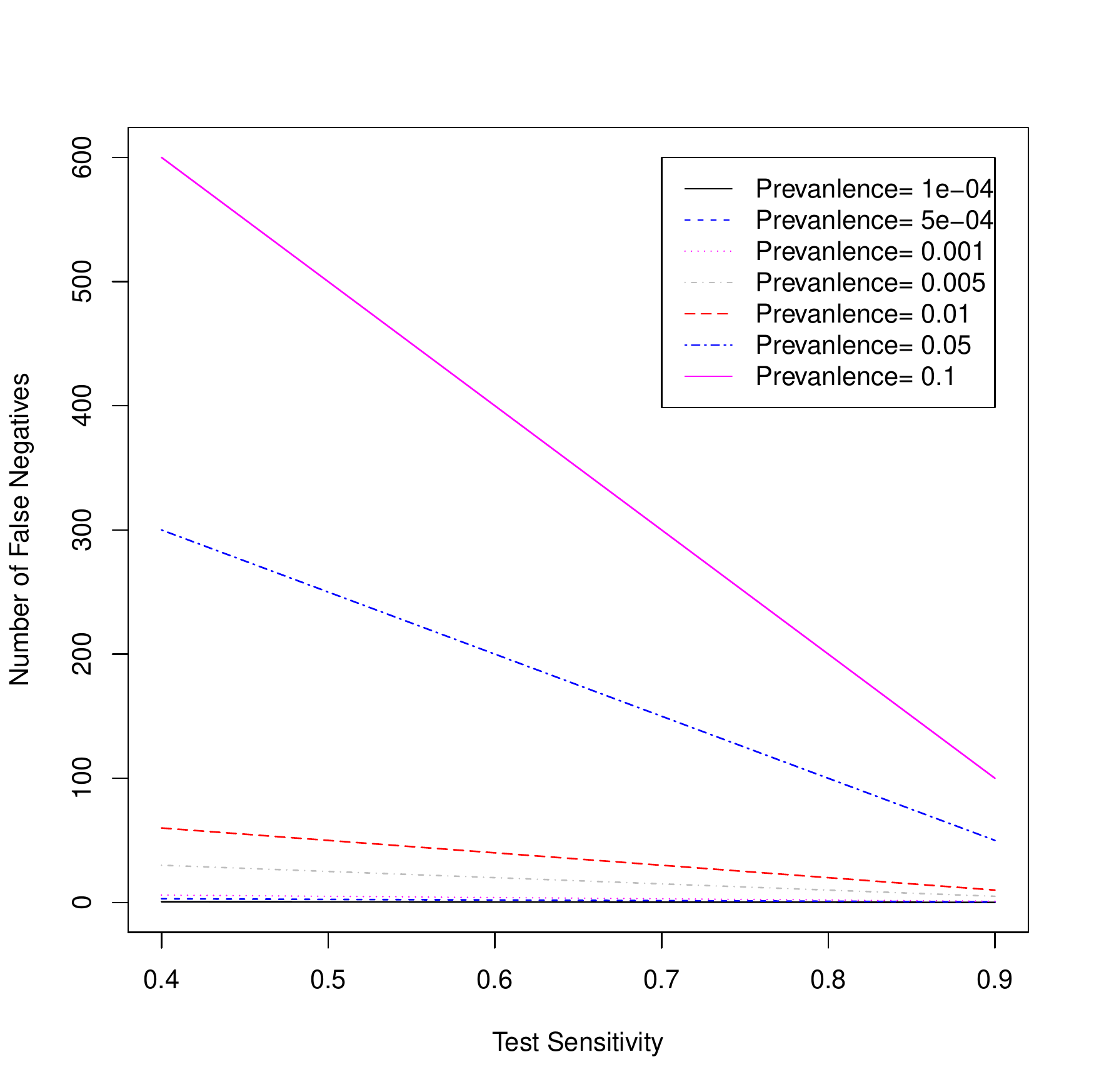}
\end{center}
{\em Figure 1: The false negative number versus the sensitivity of the test for populations of the common size 10,000 but different prevalence }

\subsection{Testing for COVID-19}

We now visualize the relationship between the false positive number ${\rm \#_{fp}}(t)$
and the specificity of the test. Figure 2 shows the results for populations with a common size $N(t)=10,000$ but with
 different prevalence.
 The false positive number becomes smaller when
the specificity of the test becomes higher. Interestingly,
the decreasing rate appears fairly stable regardless of the
prevalence value, though those drop rates are not identical.
For a given test, the false positive number is bigger for a population with a smaller prevalence, and the difference between two populations with different prevalence tends to be negligible, especially for those tests with a high specificity. In a population with 10,000 people, the false positive number can be as high as 6,000 if a test of the  specificity around 40\%  is applied to everyone in the population;  the false positive number can be lowered to about 1,000 if everyone in the population is tested with a procedure having a high specificity (such as 95\%).
Given that even for the best scenario  the false positive number is still around 1,000 for a population of size 10,000, we conclude that it is not sensible to test everyone in the population for a disease without discretion  (even if it is affordable in terms of availability of resources and the cost).\\

\noindent
{\bf Recommendation 2:} {\em Do not test everyone for COVID-19 without discretion.}\\

To further illustrate this, we consider two examples.
The accuracy of the current COVID-19 tests is not precisely
known. Based on the test performance in China and the
performance of the influenza tests,
Hutchison (2020) suggested that the sensitivity and specificity
of COVID-19 tests were estimated to be 60\% and 90\%, respectively.\\

\begin{center}
\includegraphics[width=3in]{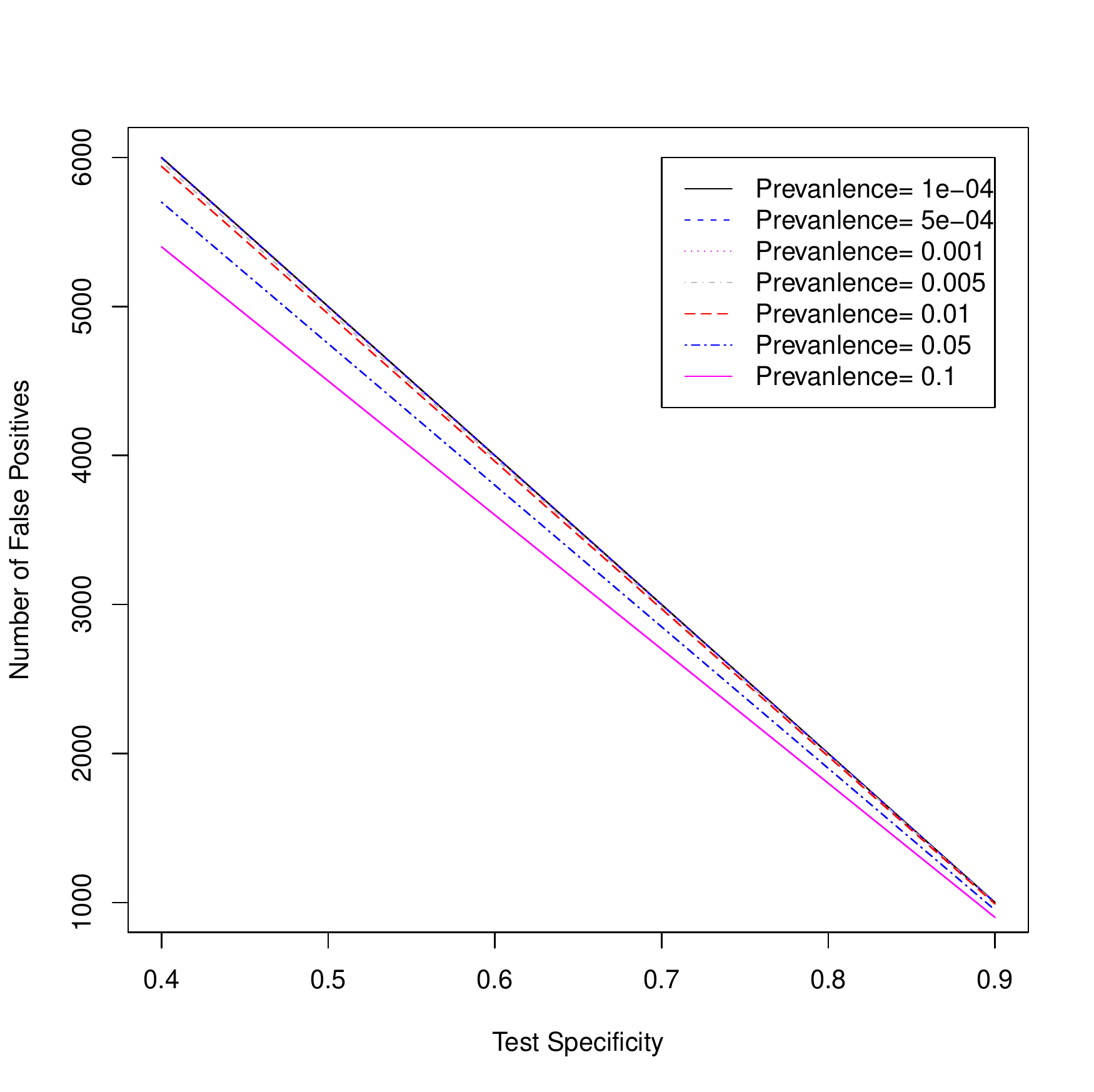}
\end{center}
{\em Figure 2: The false positive number versus the specificity of a test for populations with the common size 10,000 but different prevalence}\\

\noindent
 {\bf Example 1:} As of March 16, 2020, there were 8 confirmed cases in Waterloo Region, Ontario, Canada (CBC News 2020), whose population size is about 601,220  (Region of Waterloo 2019). That is, $P(t) \approx 0.00099797\%$ for $t$ representing
the day  March 16, 2020.
If we would test everyone in Waterloo for COVID-19, then we would expect
$$
{\rm \#_{fp}}(t) = (601,220-8)\times (1-90\%)=60,121.2
$$
and
$$
{\rm \#_{fn}}(t) = 8 \times (1 - 60\%)=3.2.
$$
That is,  three infected person would be missed, and 60,121 healthy people (i.e., near 10\% of the population in Waterloo Region) would be misdiagnosed as infected with COVID-19.
This clearly demonstrates the blunder of testing everyone in a sizable population  for COVID-19 without discretion.\\

\noindent
{\bf Example 2:} This example examines an opposite situation where the population is defined
to be a cohort of a small size.
In the period of March 11-16, 2020, among those in-person assessments and virtual visits in the clinic of Dr. Yu (the last co-author),
18 patients had flu-like symptoms and they all expressed interest  to be tested for COVID-19.\\

Assume that this small cohort has the same prevalence as that of the Waterloo Region.  If all those patients were to be tested for COVID-19,
we would then expect
$$
{\rm \#_{fp}}(t) = 18 \times (1-0.00099797\%) \times (1-90\%) = 1.78
$$
and
$$
{\rm \#_{fn}}(t) =  18 \times 0.00099797\% \times (1 - 60\%) \approx 0,
$$
where $t$ represents the short time interval March 11-16, 2020.
That is, almost no infected patients would be mis-identified  but  about 1 or 2  healthy people would be misdiagnosed as infected cases, if all  18 patients are tested for COVID-19 without being screened.\\

In fact, for those 18 patients with flu-like symptoms,
only one patient was offered testing based on the public health definition of PUI. If we perceive  that this cohort should have a higher prevalence than that of the general population in the Waterloo Region and assume that $P(t)=1/18$, then we would  expect
$$
{\rm \#_{fp}}(t) = 18 \times (1-1/18) \times (1-90\%) = 1.70
$$
and
$$
{\rm \#_{fn}}(t) =  18 \times (1/18) \times (1 - 60\%) = 0.4
$$
if everyone in this cohort would be tested for COVID-19.\\

\section{Am I Infected with COVID-19  if I have a Positive Result after Several Consecutive Negative Results?}

Since no medical tests can produce 100\% accurate results,
 both false negative and false positive results are possible when
  testing suspected patients.
  We are interested in  whether repeating the test can help improve the accuracy of the diagnosis.
   In particular, we evaluate the chance
    that a tested subject is an infected
  case, given that the first $(k-1)$ consecutive tests are negative but  the $k$th test is positive, where $k \ge 1$.
We hope to study whether it is necessary to continuously repeat the test, if consecutive negative results have been obtained after  certain repetitions. When should we stop testing in order to not miss infected cases?\\

To this end, let $Y_k^*$ represent the $k$th test result of applying the test to an individual, where $k$ is a positive integer. This binary random variable has the same distribution as that of $Y$.
We want to find the value of $k$ so that the conditional probability
$P(Y=1|Y^*_1=0, \ldots, Y^*_{k-1}=0, Y^*_k=1)$ is smaller than a tolerance value, where $k= 1, 2, \ldots$.\\

To find how the performance of the test comes into play, we express the conditional probability
$P(Y=1|Y^*_1=0, \ldots, Y^*_{k-1}=0, Y^*_k=1)$
using the sensitivity and specificity of the test as well as the prevalence.
Assuming that the test is independently applied to an individual $k$ times,
Using the Bayesian theorem gives that
\begin{eqnarray}
\nonumber && P(Y=1|Y^*_1=0, \ldots, Y^*_{k-1}=0, Y^*_k=1)\\
\nonumber &=& \frac{P(Y^*_1=0, \ldots, Y^*_{k-1}=0, Y^*_k=1|Y=1) P(Y=1)}
   {\sum_{r=0,1} P(Y^*_1=0, \ldots, Y^*_{k-1}=0, Y^*_k=1|Y=r) P(Y=r)}\\
\nonumber
&=& \frac{(1-p_{sen})^{k-1}p_{sen}P(t)}{(1-p_{sen})^{k-1}p_{sen}P(t)+p_{spe}^{k-1}(1-p_{spe})(1-P(t))}\\
&=& \frac{1}
    {1+\left(\frac{p_{spe}}{1-p_{sen}}\right)^{k-1}
       \left(\frac{1-p_{spe}}{p_{sen}}\right)
       \left(\frac{1-P(t)}{P(t)} \right)}.
\label{true-positive}
\end{eqnarray}
for $k=1, 2,  \ldots$. \\

 For any test with the specificity higher than the false negative rate,
 (\ref{true-positive}) shows that a larger value $k$ suggests a smaller  chance for
 an individual to be infected if the first positive test result appears at the $k$th test, no matter what the sensitivity of the test and the population prevalence are.
However, for a smaller number of $k$, the  sensitivity of the test and the population prevalence
do matter for quantifying the probability. \\

In  Figure 3,
 we report the conditional probability $P(Y=1|Y^*_1=0, \ldots, Y^*_{k-1}=0, Y^*_k=1)$ versus the sensitivity and specificity of the test
for   populations with different prevalence.
In  the top panel of Figure 3, we display the results for $k=1$.
As long as the test has a high specificity,
applying
 the test  to an individual
 from the population with a high prevalence  can be very informative. If the test result is positive, then it is highly likely that this individual is  infected  with the disease; this is  true, even if the test does not have a high sensitivity. However, for a population with a low prevalence (e.g., 1\%), even if the test has both a high specificity and sensitivity, the probability of correctly confirming an infected case is very low if the test is applied only once. This finding further suggests that it is unwise to apply
 a test to everyone from the population with a low prevalence. It is advised that prior measures should be taken to identify a suspected subpopulation (or a group of presumptive cases) so that the resulting prevalence becomes high. Then applying the test to individuals in this subpopulation can increase accuracy to identify infected cases.\\

The bottom panel of Figure 3
shows the results for $k=3$.
 For  a test with a low sensitivity,
there is a high chance that a patient is infected even when the first positive result appears at the third test,
if we test individuals from the population with a high prevalence (such as 85\%).
However, if the population has a small prevalence (such as 1\%), it is unlikely  that the  patient with the first positive result occurring at the third try is infected.\\

\noindent{\bf Recommendation 3:} {\em With a given test,
  when interpreting a positive result after consecutive negative results, caution should be taken for patients coming from
different cohorts with different prevalence.}\\

\begin{center}
\includegraphics[width=6.5in]{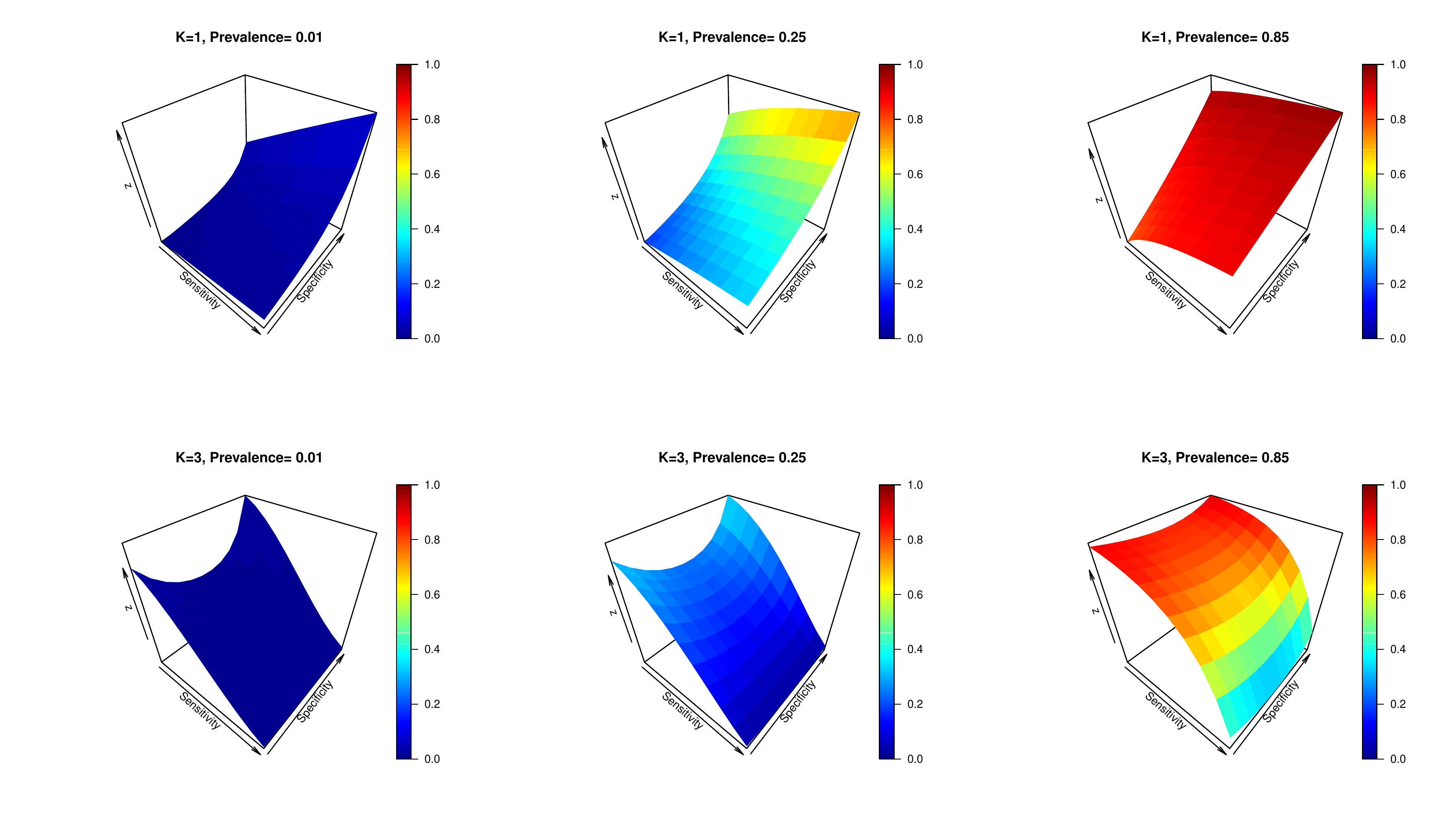}
\end{center}
{\em Figure 3: The conditional probability for confirming a case by repeating the test  $k$ times versus the sensitivity and specificity of the test: $k=1, 3$ and the prevalence $P(t)=0.1, 0.25, 0.85$; The color shows the magnitude of the probability.}\\

To further visualize how the cohort prevalence affects  the probability of identifying an infected case, given consecutive negative results followed by  a positive result,
 we examine
 the COVID-19 test described by Hutchison (2020) as opposed to
 the {\em COVID-19 IgM/IgG Rapid Test}, a test newly released by the ISO13485 registered company {\em BioMedomics}.
On March 8, 2020, the company
announced that it has received CE Mark-IVD certification for its new
 test to help diagnose novel
COVID-19.
This test, available only for research use  at this stage,
takes
  15 minutes to obtain the result and
can be used for rapid screening of COVID-19 carriers who are symptomatic or asymptomatic.
The sensitivity and specificity of the test
were  estimated to be 88.66\% and 90.63\%, respectively, based on  the test results for
   525 infected cases and 128 non-SARS-CoV-2 infection patients (BioMedomics 2020). \\

We  graph the results in Figure 4.
There is a high chance that the tested person contracts COVID-19 if the  result of applying the COVID-19 IgM/IgG Rapid Test
is positive at the first test, unless the population prevalence is very small; the chance is higher for testing patients coming from a cohort with a higher prevalence. For the COVID-19 IgM/IgG Rapid Test, if the positive result
occurs only at the 4th test, then the chance of the tested subject is infected is very slim unless the patient comes from a cohort with a high prevalence (such as 60\% or higher); if the first positive result
occurs at the 6th test,
the chance of the tested subject is infected is almost 0 no matter which cohort this patient comes from.
However, for the COVID-19 test described by Hutchison (2020), even   the first five test  results are negative,
 there is still a good
chance that the tested subject is infected with COVID-19. \\

\noindent{\bf Recommendation 4:} {\em  With the same cohort of patients, different interpretations should be given for the positive result after the same number of consecutive negative results that are obtained from different tests with different sensitivities and specificities.}\\

%
%

\begin{center}
\includegraphics[width=5.5in]{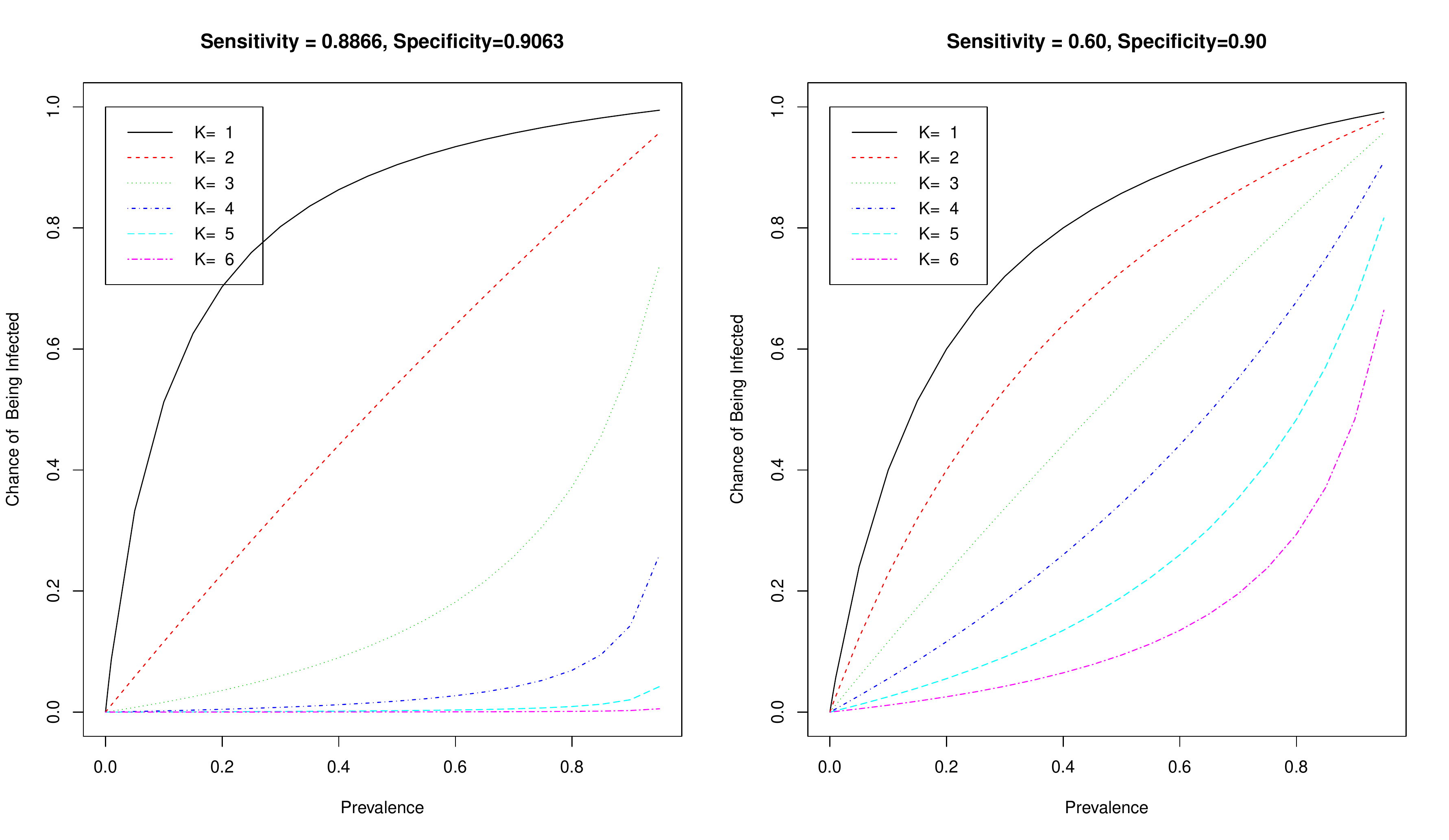}
\end{center}
{\em Figure 4: The probability of identifying an infected case after $k-1$ consecutive negative results
followed by a positive result: the plots of the  probability versus the prevalence for $k=1, \ldots, 6$; The left panel is for
the COVID-19 IgM/IgG Rapid Test and the right panel is for the COVID-19 test described by Hutchison (2020).}\\

\section{Can I Get COVID-19 Twice?}

As of March 13, 2020,
in mainland China, there have been more than 100 reported cases of cured patients released from hospitals, who later tested positive for COVID-19 a second time. In China's Guangdong province,  14\%  of people who recovered in the province were later retested to be positive.
Similar cases have been reported in Japan and South Korea (Guzman 2020). This prompts the question
``Can I get COVID-19 twice?" Put in other words, we want to know whether we can
 differentiate the two scenarios: (1) patients who contracted COVID-19 again after they were cured,
and (2) patients who have never really been cured and their discharge was due to the
false negative test results.\\

Researchers perceive that reinfection is an unlikely explanation for patients who test positive a second time.
Testing errors and releasing patients from hospitals  prematurely are  very likely the reason for reports of patients who retest positive.
Although it has not been proved that people who have contracted COVID-19 are immune, this is very likely the case, as noted by Anthony Fauci, Director of the National Institute of Allergy and Infectious Disease (Guzman 2020). \\

To assess this perception,
we evaluate the chance of incorrectly claiming  a COVID-19 carrier to be cured after obtaining several consecutive negative results.
 We are interested in assessing the conditional probability
$P(Y=1|Y^*_1=0, \ldots, Y^*_{k-1}=0, Y^*_k=0)$ for $k=1, 2, \ldots$.
Assuming that the test is independently applied to a patient $k$ times, then
using the Bayesian theorem, we obtain  that
\begin{eqnarray}
\nonumber && P(Y=1|Y^*_1=0, \ldots, Y^*_{k-1}=0, Y^*_k=0)\\
\nonumber &=& \frac{P(Y^*_1=0, \ldots, Y^*_{k-1}=0, Y^*_k=0|Y=1) P(Y=1)}
   {\sum_{r=0,1} P(Y^*_1=0, \ldots, Y^*_{k-1}=0, Y^*_k=0|Y=r) P(Y=r)}\\
&=& \frac{(1-p_{sen})^{k}P(t)}{(1-p_{sen})^{k} P(t)+p_{spe}^{k}\{1-P(t)\} }
\label{fale-negative}
\end{eqnarray}
for $k=1, 2, \ldots$. \\

In Figure 5, for  populations with different prevalence,
 3-dimensional graphs were made to show how the conditional probability for missing an infected  case
 with $k$ consecutive negative results may depend on the sensitivity and specificity of the test, where $k=1,3$.
 It is clearly seen that if the test has a low sensitivity, the chance of missing an infected case is  high if the test is done only once for populations with a large prevalence, even if the specificity of the test is high; the larger the prevalence, the higher the chance of missing.
 For a test with a reasonably large sensitivity,
 the more we test, the smaller the chance of missing an infected case.\\

More rigorously,
we rewrite the conditional probability
 (\ref{fale-negative}) as
 \begin{equation}
 P(Y=1|Y^*_1=0, \ldots, Y^*_{k-1}=0, Y^*_k=0)
= \frac{1}{1+ \left( \frac{p_{spe}}{1-p_{sen}}\right)^{k}\times \left( \frac{1- P(t)}{P(t)} \right)}
\label{fale-negative1}
\end{equation}
for $k=1, 2, \ldots$.
The probability  (\ref{fale-negative1}) decreases as $k$ increases {\em if and only if}
$$
\frac{p_{spe}}{1-p_{sen}} > 1,
$$
which is satisfied by a test
with the sensitivity larger than the false positive rate (i.e., $1-p_{spe}$)
or the specificity  higher than the false negative rate (i.e., $1-p_{sen}$).
This condition must be met by any test in use, otherwise, there is no point of
using a test that is even worse than a random guess. Hence, for any test in use, increasing the test number reduces the chance of mistakenly discharging infected patients.\\

\noindent
{\bf Recommendation 5:} {\em To reduce the chance of missing a COVID-19 carrier
 based on negative test results, it is important to increase the number of tests.}\\

\begin{center}
\includegraphics[width=6.5in]{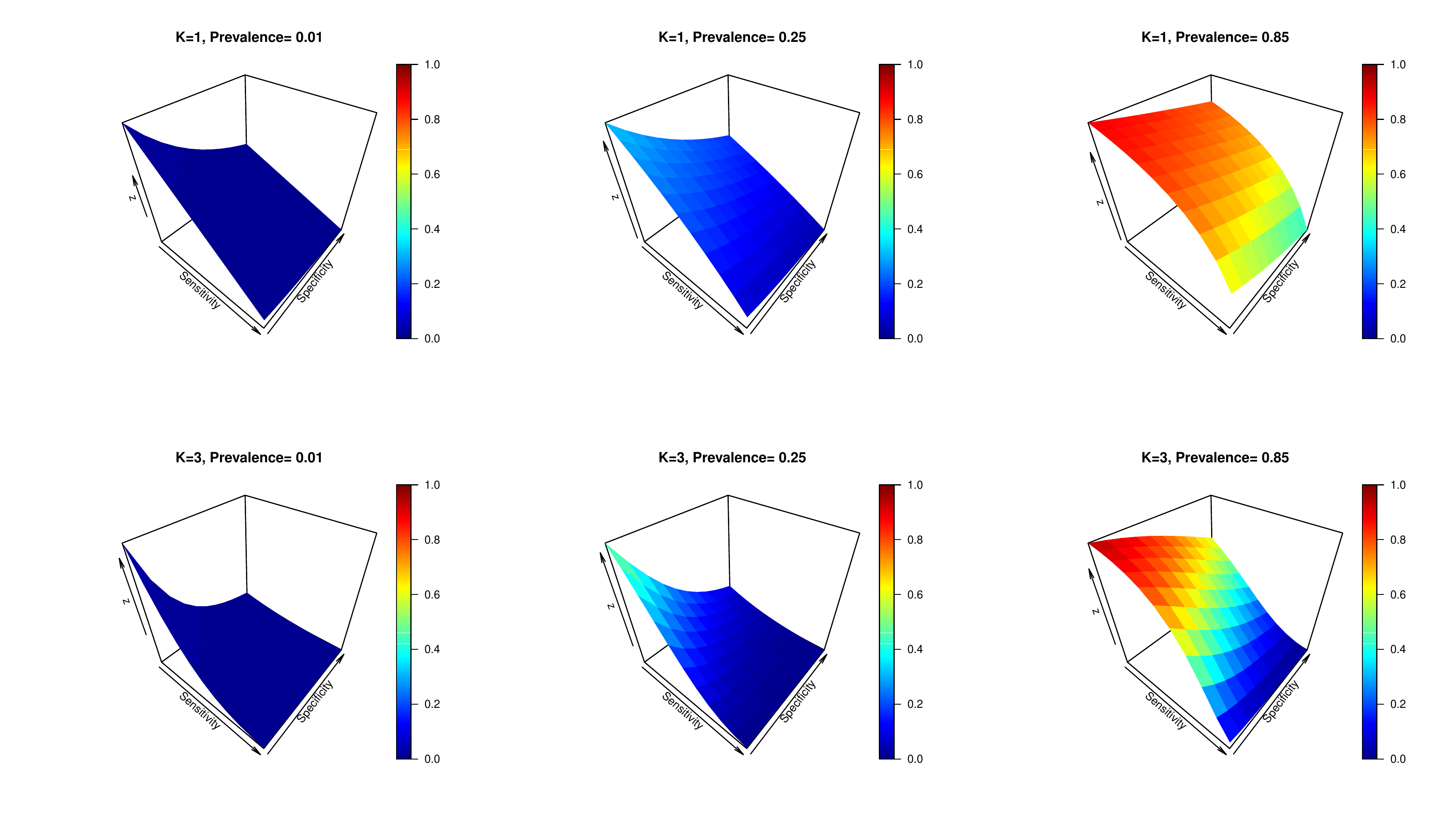}
\end{center}
{\em Figure 5: The chance of missing a case with $k$ consecutive negative results versus the sensitivity and specificity of the test:
$k=1, 3$ and the prevalence $P(t)=0.01, 0.25, 0.85$; The color shows the magnitude of the chance.}\\

To see how doctors may implement this recommendation when they need to discharge inpatients with COVID-19-like
symptoms, we compare
 the COVID-19 test described by Hutchison (2020) to
 the {\em COVID-19 IgM/IgG Rapid Test}.
Figure 6 displays how the probability of missing an infected case after receiving
$k$ consecutive negative test results depends on the population prevalence of COVID-19 for
 $k=1, 2, \ldots, 6$.
With the COVID-19 IgM/IgG Rapid Test, having two consecutive negative test results ensures  a nearly zero chance of missing infected cases if testing patients admitted to the ward with less than 50\%  COVID-19 carriers; for the inpatient ward with about 80\%  COVID-19 carriers,  obtaining 3 consecutive negative test results warrants a slim chance of missing infected cases; receiving  4 consecutive negative results  is enough for discharging any inpatients.\\

In comparison, the COVID-19 test described by Hutchison (2020) has  about 20\% smaller sensitivity than the COVID-19 IgM/IgG Rapid Test. It requires  more consecutive negative results than the COVID-19 IgM/IgG Rapid Test for retaining a slim chance of missing infected cases, and  different numbers of consecutive negative results produced by this test yield more different
 probabilities of missing infected cases than those obtained from the COVID-19 IgM/IgG Rapid Test. This comparison also illustrates how a high sensitivity of a test can make a difference in reducing the chance
 of missing infected cases based on consecutive negative results.\\

\noindent
{\bf Recommendation 6}: {\em Different numbers of consecutive negative test results are required
to discharge inpatients admitted to  wards with different prevalence.
For the  COVID-19 test described by Hutchison (2020), to ensure the chance of mis-discharging to be smaller than 5\%,
2, 3, 4, 5, and 6  consecutive negative test results are needed to discharge inpatients in a ward with the
prevalence about 20\%, 40\%, 50\%, 70\%, and 80\%, respectively.}\\

\begin{center}
\includegraphics[width=5.5in]{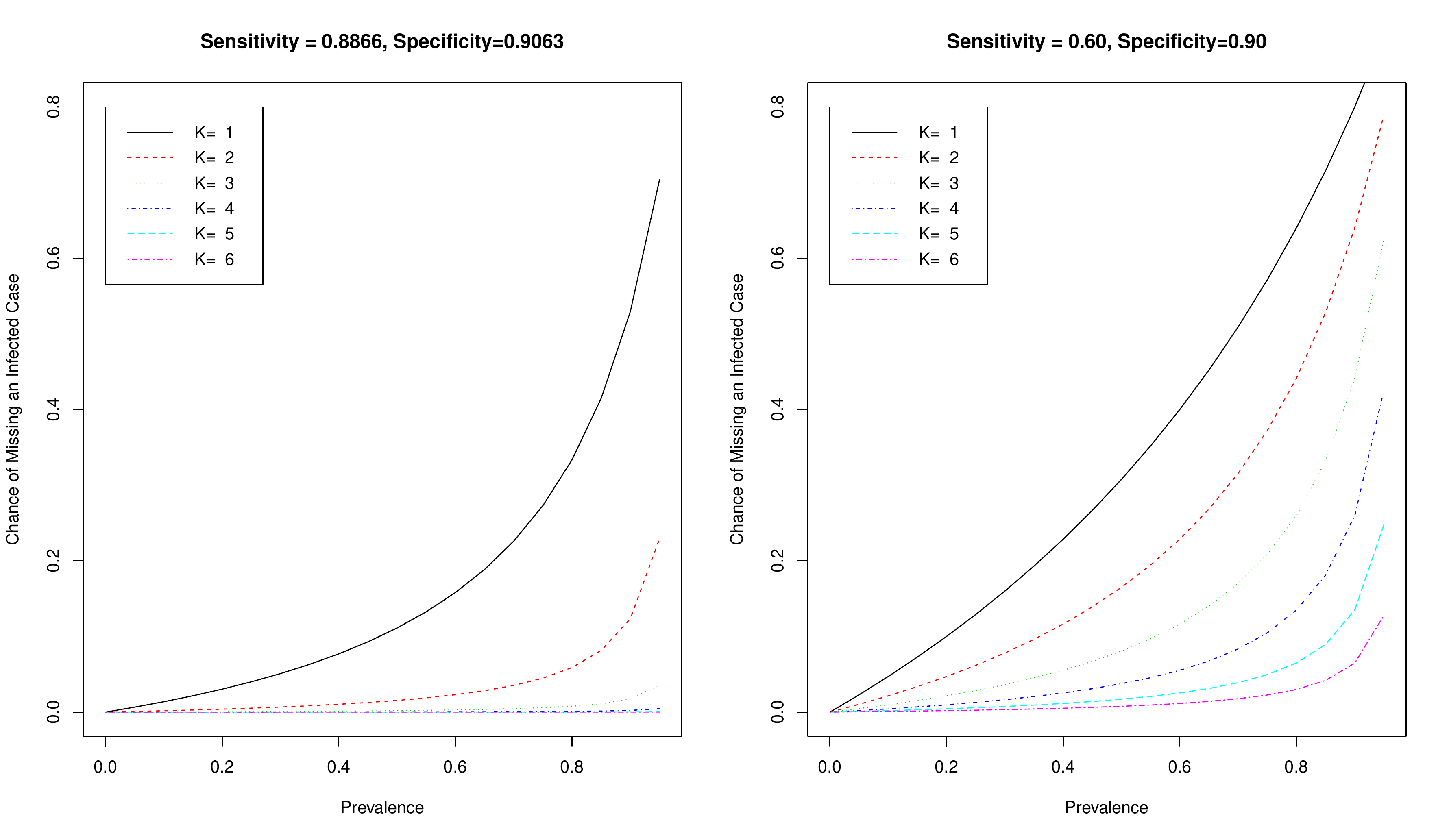}
\end{center}
{\em Figure 6: The conditional probability for missing an infected case based on $k$ consecutive negative results: the plots of the conditional probability versus the prevalence for $k=1, \ldots, 6$; The left panel is for
the COVID-19 IgM/IgG Rapid Test and the right panel is for the COVID-19 test described by Hutchison (2020)}\\

\section{Analysis of Diamond Princess Data}\label{aodpd}

To illustrate our discussion, we analyze the data of the Diamond Princess cruise for the period of January 19, 2020 (the day before  {\em patient zero} embarked on the cruise) to February 20, 2020 when all passengers were disembarked (Princess 2020).
Diamond Princess was on a 14-day round trip itinerary,  departing from Yokohama in Tokyo on January 20, 2020 and  returning on February 4, 2020. There were 2,666 guests and 1,045 crew on board. {\em Patient zero}
traveled for five days on Diamond Princess and disembarked in Hong Kong on January 25. During those five days, patient zero did not report being ill; he was tested positive for COVID-19 on February 1, six days after leaving the ship.\\

On February 4, 10 people were tested positive for COVID-19 among the first batch of tested passengers. In subsequent days, more guests were tested positive. People with positive test results were  transported to local hospitals for medical care.
Table 1 displays the number of people whose test results were positive on different days. Using the notation in Section \ref{naf}, January 19 is taken at $t=0$ on which $N(0)=N_h(0)=3711$ and $N_s(0)=0$;  January 20 is taken as $t=1$ on which  $N(1)=N(0)$ and $N_s(1)=1$; on February 4 (i.e., $t=17$), $N_s(17)=10$ and $N(17)=N(0)-10$. The left panel of Figure 8 displays the day-dependent prevalence, and in the middle panel we report the changing population size using the red curve.\\

Without discretion, we would use the COVID-19 test described  by Hutchison (2020) to test {\em all} passengers  {\em every day} starting $t=17$, then the daily numbers of false positive and negative results can be worked out by (\ref{false-number}). We visually display those numbers in the middle panel (blue dashed curve) and the right panel in Figure 7. Clearly, when $P(t)$ is very small, $\#_{fp}(t)$ is close to $N(t)$ and $\#_{fn}(t)$ is near 0; when $P(t)$ becomes larger, $\#_{fp}(t)$ becomes smaller but $\#_{fn}(t)$ gets larger.\\

Though
having a large value of $\#_{fp}(t)$ would not exacerbate the virus spread, it would  waste limited medical resources. Having a nonzero value of
$\#_{fn}(t)$ would be harmful because those undetected infected cases would be spreaders of the virus; on the day $t=30$ of disembarking all the passengers, there could be 300 missed COVID-19 carriers if we  test
{\em everyone} on the ship without carefully screening.\\

It is sensible to focus on testing suspected patients to reach a balance between the use of limited medical resources and the result accuracy. By (\ref{true-positive}), the COVID-19 test should be repeated at
least three times for a suspected patient, and all three negative results would ensure the chance of missing
an infected case to be under 5\%.\\

\begin{center}
\begin{table}
\tabcolsep=4pt
\caption{COIVD-19 Data from Diamond Princess Cruise}
\begin{tabular}{l c c  c c  c c  c c  c c  c c  c c  c c  c c  } \\ \hline
Day&1-16&17&18&19&20&21&22&23&24&25&26&27&28&29&30&31&32&33-44 \\ \hline
\# Cases& 0&10&10&41&3&0&6&65&39&0&47&0&134&0&99&88&79&84 \\ \hline
\end{tabular}
\end{table}
\end{center}

\begin{center}
\includegraphics[width=4in]{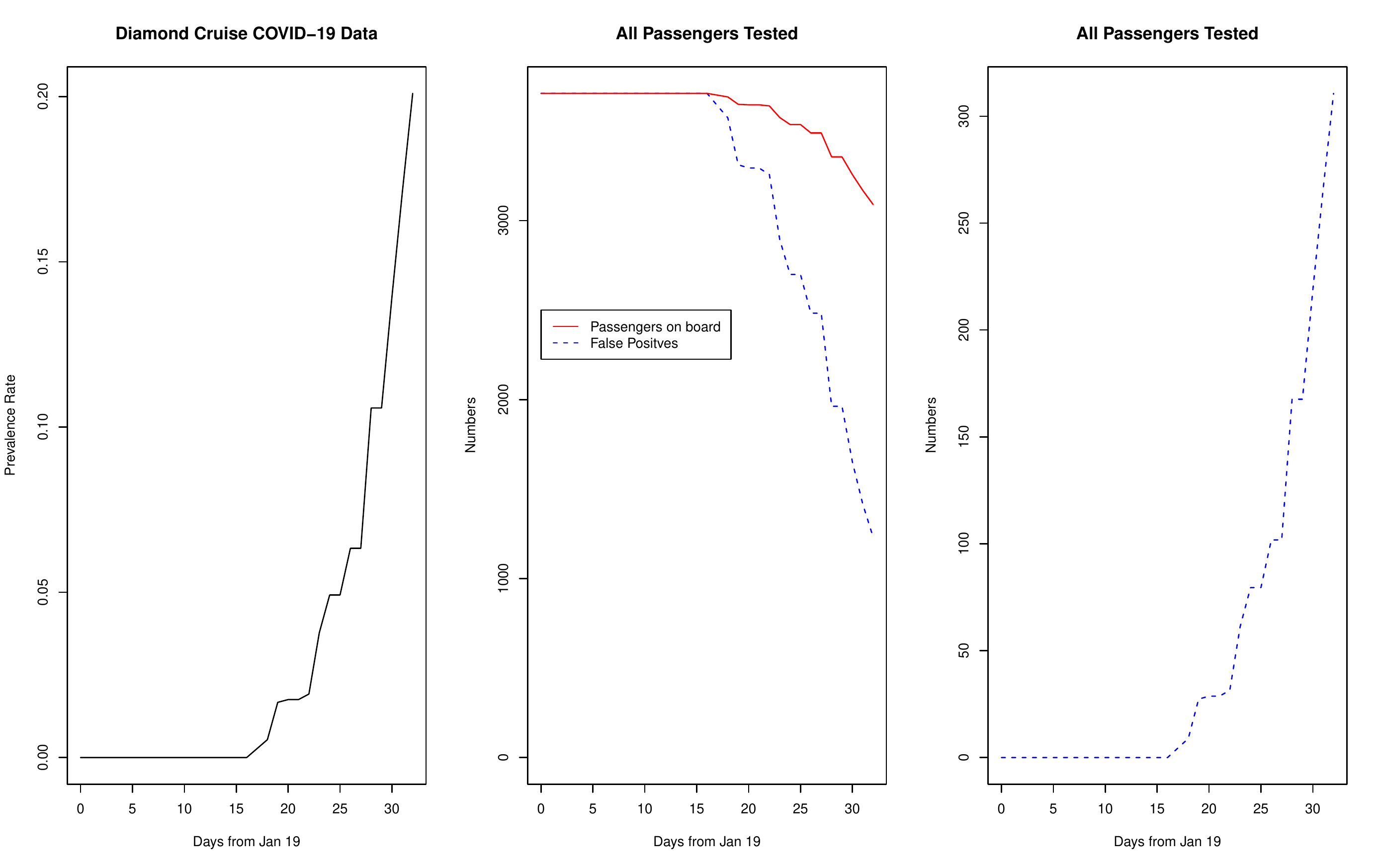}
\end{center}
{\em Figure 7: Analysis of the Diamond Princess data if everyone would be tested everyday from day 17 to day 44: the left panel records the prevalence versus the date, the right panel shows the false negative number versus the date, and
 the middle panel displays the  false positive number  versus the date (in blue) and
 the population size versus the date (in red).}

\section{Conclusion and Discussion}

In this article we take the statistical standpoint and examine several aspects of COVID-19 testing.
We evaluate the uncertainty induced from the imperfectness of  COVID-19 tests
and make recommendations. In summary, we highlight the following points:\\

\begin{enumerate}

\item [1.] Testing everyone {\em without discretion} is not recommended. A large number of false
negatives is typical for medical tests with a low sensitivity.
A large number of false positives is produced from a test with a lower
specificity. It is not  feasible to test everyone who shows flu-like symptoms.
It is important to prioritize the testing of people who need it the most,
not only for the economical considerations of the availability of test kits, but also
for the statistical concerns of controlling false positive or negative results.\\

Even with enough test kits for the entire population, it is advised to identify
suitable candidates for the test. It is unwise to test {\em everyone}
in the population without careful discretion.
Medical personnel generally believe that
people in the following groups should be prioritized to test for COVID-19 (Ferran 2020):
(1) those at high risk such as health care workers who have been in contact with COVID-19 patients, (2) symptomatic people in areas with high infection rates,
(3) people 65 years of age and older with chronic health issues, such as heart disease, lung disease, and diabetes, and (4) patients who have immunodeficiency diseases.\\

\item [2.] The accuracy of the COVID-19 test described by Hutchison (2020) is not precisely
known. Based on the test performance in China and the
performance of the influenza tests,
the test is estimated
to correctly identify
around 60\% of the patients with the disease and correctly identify 90\%
of the patients that are disease-free.
The low sensitivity of
the current tests for COVID -19
would yield a large number of false negative if we test everyone in a cohort with a low
prevalence. That is,
 a large number of people having the virus
would be misdiagnosed as healthy if we attempt to test the entire cohort which has not been carefully scrutinized for COVID-19.\\

\item [3.] The tests should only be applied to presumptive
people in order to control false positive and false
negative results.
 With limited resources, targeting the presumptive
people and applying the tests {\em repeatedly} is
the best strategy to make sure the true COVID-19 carriers are not missed and discharged inpatients are truly virus-free.
\end{enumerate}

\vspace{3mm}

Our discussion is carried out exchangeably at the {\em population} level and on
the basis of an {\em individual}. When
interpreting
the results, caution should be taken for this difference.
In the initial stage, COVID-19 was transmitted in an unrecognized way. Limited testing capacity and strict testing  criteria delayed the identification of COVID-19 carriers in many countries (Sullivan 2020). Our discussion here is useful to
help make prudent decisions from the {\em administrative
standpoint} to optimize the usage of limited resources of test kits, healthcare workers, and medical facilities.
However,  from the {\em perspective of an individual} thinking ``better safe than  sorry",
one might argue that having
 a false positive result is tolerable but   a
 false negative is troublesome. From  a  physician's viewpoint,  missing
 an infected case can be more detrimental than misdiagnosing a COVID-19-free patient as a carrier. Such an error
 would cause a spread of virus and delay medical care  for the COVID-19 infected patients.
This article provides the assessment as to how likely we may miss identifying COVID-19 carriers based on consecutive negative results and how many times we should test a suspected COVID-19 patient to reduce the chance of errors. On equal footing, our discussion provides the guidelines for discharging inpatients who are treated for COVID-19. \\

There are several limitations of our discussion.
In the discussion of the Diamond Princess Cruise Data in Section
\ref{aodpd}, we ignore the fact that the population size $N(t)$ and the prevalence $P(t)$ are error-contaminated. In our discussion, the
dynamic number of COVID-19 carriers
 $N_s(t)$ is taken as  the confirmed  cases for each day $t$.
The true value of $N_s(t)$ is, however, highly likely to be larger than the reported number for  day $t$   since some infected passengers were  asymptomatic and thus were not being tested on day $t$ for $t=17, 18, \ldots, 34$.\\

While we consider
a dynamic framework in Section \ref{naf} to characterize
the change in the population size and the relationship between
the number of infected people and the number of healthy people, our discussion on the COVID-19 status for individuals focuses on a static state to highlight the ideas.
 More specifically, time-dependent status $Y(t)$, instead of $Y$, should be used to reflect
the time-dependent status for individuals in the population on day $t$.
As our understanding of COVID-19 grows and
more accurate test kits become available, our development
should be modified and the  recommendations can be made by incorporating
time windows to reflect influencing  factors, including the evolving stage of COVID-19, preventive
measures done by the local administration and the government,
and the changes in the public social behavior for different periods.\\

Another notable issue is that our discussion merely  investigates
testing errors from the statistical angle to quantify the randomness
and uncertainty associated with the test inaccuracy. We have not explicitly accommodated in the development
the characteristics of patients, such as
age, severity of COVID-19-like symptoms, health conditions, and the medical history. This is caused by
 the lack of information on the sensitivity and specificity
of available test kits estimated from different  cohorts of patients with varying medical conditions.
As a last recommendation, we  suggest that rather than reporting
an {\em overall} sensitivity and specificity for a developed test,
developers of COVID-19 tests should
evaluate a sequence of the sensitivities and specificities of the test applicable
to different cohorts of patients.  This will allow physicians and medical personnel
to make more precise decisions to accommodate the personalized-features of the patients.\\

\noindent{\bf Recommendation 7:} {\em Instead of being assessed by an
{\em overall} sensitivity and specificity,
 the performance of COVID-19 tests should be evaluated in a more refined measure by
reporting their  sensitivities and specificities obtained from
 the stratified population by the patient's medical conditions.}\\

In the article, we compare two COVID-19 tests, the COVID-19 test described by Hutchison (2020) and
the COVID-19 IgM-IgG Rapid Test. While the calculations show that the COVID-19 IgM-IgG Rapid Test outperforms
the COVID-19 test described by Hutchison (2020), we are not ready to recommend to replace
the latter test  by the former one. While the sensitivity
and specificity of the COVID-19 IgM-IgG Rapid Test are higher than those of the COVID-19 test described by Hutchison (2020), these results are obtained from different groups of patients whose conditions
may differ considerably and the sizes may not be comparable either.\\

Having fast and effective  test tools is crucial for controlling the rapidly evolving COVID-19 course while
emerging research results offer new ways for testing COVID-19.
For instance,
investigating  the temporal changes of COVID-19 pneumonia on CT scans,
Shi et al. (2020) found that
a CT could  be a useful tool to
 detect COVID-19 pneumonia, even for asymptomatic individuals.
Their findings
suggested that CT scans can  be considered
as a screening tool together with RT-PCR for
patients who  traveled recently  or have had close
contact with an infected individual. Furthermore, CT scans may
 be an important screening tool in the
small proportion of patients who have false-negative
RT-PCR results (Lee, Ng and Khong 2020). Announced on March 21, 2020,
diagnostics company Cepheid received emergency authorization from the U.S. FDA to use its rapid molecular test, {\em SAR-CoV-2 Xpert Xpress},  that can detect     COVID-19 in 45 minutes for point-of-care patients (Scipioni 2020). A
review of the current
laboratory methods available for testing COVID-19 was given by Loeffelholz and Tang (2020).\\

Our discussion here considers repetitions of  the same test procedures.
With multiple test kits becoming available, we face the decision of choosing suitable  test kits to reach a balance among several key factors. This includes but not limited to the time of acquiring results, the associated cost,
  the test accuracy, and the suitability for
patients with
different conditions.
It is useful
to ponder the question: How do we use them effectively? We may apply a fast but less accurate test kit to do screening, and then apply a more accurate but time-consuming and costly test to do further checks.
 If repeated tests need to be done, we need to decide the number of
 the tests   in order to not miss the infected patients.
  In addition,
in our discussion of repeating the test for COVID-19, we have not looked into the issue of the gap time between two consecutive tests.

\section*{Acknowledgements}

The research of Yi and He is partially supported by fundings from the Natural
Sciences and Engineering Research Council of Canada (NSERC).
Yi is Canada
Research Chair in Data Science (Tier 1). Her research was undertaken, in part, thanks
to funding from the Canada Research Chairs Program.


\bibliographystyle{unsrt}

\begin{thebibliography}{99}

\bibitem{}
BioMedomics (2020). BioMedomics receives CE-IVD certification for its new COVID-19 IgM-IgG Rapid Test for novel coronavirus. https://www.biomedomics.com
/products/infectious
-disease/covid-19-rt/?from=singlemessage\&isappinstalled=0.

\bibitem{}
CBC News (2020). 8 cases of COVID-19 identified in Waterloo region. March 16th, 2020. https://www.cbc.ca/news/canada/kitchener-waterloo.


\bibitem{}
Ferran, M. (2020). Should I be tested for Coronavirus? Here's what you need to know about COVID-19 tests.
{\em The Conversation}, March 13, 2020. https://theconversation.com
/how-does-the-coronavirus-test-work-5-questions-answered-133118.

\bibitem{}
Guzman, J. (2020). Can you get coronavirus twice? https://thehill.com/changing-america/well-being/prevention-cures/487436-can-you-get-coronavirus-twice.

\bibitem{}
Hutchison, R. L. (2020). The accuracy of COVID-19 tests.
https://www.kevinmd.com
/blog/2020/03/the-accuracy-of-covid-19-tests.html.

\bibitem{}
Lee, E. Y. P., Ng, M.-Y., and Khong, P. L. (2020).
COVID-19 pneumonia: what has CT taught us? {\em The Lancet Infectious Disease}.
Published online February 24, 2020 https://doi.org/10.1016/S1473-3099(20)30134-1.


\bibitem{}
Loeffelholz, M. J. and  Tang, Y.-W. (2020). Laboratory diagnosis of emerging human
coronavirus infections — The state of the art. {\em Emerging Microbes \& Infections}, https://doi.org/10.1080/22221751.2020.1745095.


\bibitem{}
Princess (2020). Diamond Princess updates.
https://www.princess.com/news/notices
\_and\_advisories/notices/diamond-princess-update.html.

\bibitem{}
Region of Waterloo (2019). Year-End 2018 Population and Household Estimates for Waterloo Region. Report: PDL-CPL-19-14. March 19, 2019.

\bibitem{}
 Scipioni, J. (2020). FDA grants `emergency use' coronavirus test that can deliver results in 45 minutes.
 https://www.cnbc.com/2020/03/21/fda-grants-emergency-use-coronavirus-test-that-can-deliver-results-in-45-minutes.html.



\bibitem{}
Sharkawy, A. (2020).
Abdu Sharkawy is on Facebook. https://m.facebook.com
/abdu.sharkawy/posts/2809958409125474.

\bibitem{}
Shi, H., Han, X., Jiang, N.,  Cao, Y.,  Alwalid, O., Gu, J.,  Fan, Y., and  Zheng, C. (2020). Radiological findings from 81 patients
with COVID-19 pneumonia in Wuhan, China: A descriptive study.
{\em The Lancet Infectious  Disease}. Published online Feb 24. https://doi.org/10.1016/
S1473-3099(20)30086-4.

\bibitem{}
Sullivan, P. (2020).
CDC testing limits delayed finding coronavirus cases, Washington officials say.
https://thehill.com/policy/healthcare/public-global-health/485311-cdc-testing-limits-delayed-
finding-coronavirus-cases?from
=singlemessage\&isappinstalled=0.

\bibitem{}
WHO Situation Report (2020).
WHO Director-General's opening remarks at the media briefing on COVID-19 - 11 March 2020.
https://www.who.int/dg/speeches/detail/who-director-general-s-opening-remarks-at-the-media
-briefing-on-covid-19---11-march-2020.


\bibitem{}
Wong, J. E. L., Leo, Y. S., Tan, C. C. huan Tan (2020).
COVID-19 in Singapore - current experience: Critical global issues that require attention and action.
 {\em Journal of American Medical Association}. Published online February 20, 2020. doi:10.1001/jama.2020.2467.


\bibitem{}
Xiao, Y. and  Torok, M. E. (2020).
Taking the right measures to control COVID-19.
{\em Lancet Infectious Disease}.
Published Online
March 5, 2020. https://doi.org/10.1016/S1473-3099(20)30152-3.

\bibitem{}
Zhou, F., Yu, T., Du, R., Fan, G., Liu, Y., Liu, Z., Xiang, J., Wang, Y., Song, B., Gu, X., Guan, L., Wei, Y.,
Li, H., Wu, X., Xu, J., Tu, S., Zhang, Y., Chen, H., and Cao, B. (2020).
Clinical course and risk factors for mortality of adult
inpatients with COVID-19 in Wuhan, China: A retrospective
cohort study. {\em The Lancet}.
Published Online, March 9, 2020.
https://doi.org/10.1016/S0140-6736(20)30566-3.



\end{thebibliography}

\end{document}